\documentclass[CEJP,PDF]{cej} 
\usepackage{layout}
\usepackage{amsmath}
\usepackage{textcomp}
\usepackage{hyperref}

\title{Numerical study of QCD phase diagram at high temperature and density by a histogram method}

\articletype{Research Article}

\author{Shinji~Ejiri\inst{1}
        \email{ejiri@muse.sc.niigata-u.ac.jp, {\rm Speaker.}},
        Sinya~Aoki\inst{2}$^,$\inst{3},
        Tetsuo~Hatsuda\inst{4}$^,$\inst{5}$^,$\inst{6},
        Kazuyuki~Kanaya\inst{2},
        Yoshiyuki~Nakagawa\inst{1},
        Hiroshi~Ohno\inst{2},
        Hana~Saito\inst{2},
        Takashi~Umeda\inst{7}
        (WHOT-QCD~Collaboration)}
\institute{
     \inst{1} Graduate School of Science and Technology, Niigata University,
              Niigata 950-2181, Japan
     \inst{2} Graduate School of Pure and Applied Sciences, University of 
              Tsukuba, Tsukuba, Ibaraki 305-8571, Japan
     \inst{3} Center for Computational Sciences, University of Tsukuba, 
              Tsukuba, Ibaraki 305-8577, Japan
     \inst{4} Department of Physics, The University of Tokyo, 
              Tokyo 113-0033, Japan
     \inst{5} IPMU, The University of Tokyo, Kashiwa 277-8583, Japan
     \inst{6} Theoretical Research Division, Nishina Center, RIKEN, 
              Wako 351-0198, Japan
     \inst{7} Graduate School of Education, Hiroshima University, 
              Hiroshima 739-8524, Japan
          }

\abstract{
We study the QCD phase structure at high temperature and density adopting a histogram method. Because the quark determinant is complex at finite density, the Monte-Carlo method cannot be applied directly. We use a reweighting method and try to solve the problems which arise in the reweighting method, i.e. the sign problem and the overlap problem. We discuss the chemical potential dependence of the probability distribution function in the heavy quark mass region and examine the applicability of the approach in the light quark region.}


\begin{document}
\maketitle


\section{Introduction}
\label{sec:intro}

It is important to study the QCD phase diagram including its quark mass dependence, as well as the temperature $(T)$ and chemical potential $(\mu)$ dependences, for understanding the nature of the QCD phase transition at high temperature and density.
The phase transition is expected to be first order when the up, down and strange quark masses are sufficiently large or small, and becomes crossover in the intermediate region between them.
We expect a second order transition in the chiral limit of 2-flavor QCD 
and the nature of the chiral transition changes as 
the strange quark mass decreases.
Moreover, the phase structure at nonzero $\mu$ is also expected to have a rich phase structure.
To study the phase diagram, the probability distribution function (histogram) of a physical quantity provides us with a useful method to determine the order of phase transitions. 
We expect that the distribution has two peaks at a first order transition when the volume is finite. 
Hence, we can identify the nature of the transition through the shape of the distribution function.
In Sec.~\ref{sec:heavy}, we study the phase structure in the $T$, $\mu$ and quark mass parameter space measuring the probability distribution function when the quark is heavy, as the first step. 
We then extend this analysis to the light quark region and discuss the applicability of our approach in Sec.~\ref{sec:light}.

\section{Plaquette effective potential in the heavy quark region}
\label{sec:heavy}

The probability distribution function for the plaquette value is defined by
\begin{eqnarray}
  w(P; \beta, \kappa, \mu) 
  = \int {\cal D} U\, \delta\left(\hat{P}[U]-P\right) \left[\det M(\kappa, \mu) \right]^{N_{\rm f}}e^{6N_{\rm site} \beta \hat{P}}.
\label{eq:pdist}
\end{eqnarray}
Here, we use the standard plaquette gauge action and $N_{\rm f}$ flavor Wilson quark action. 
$\hat{P}$ is the plaquette operator in the action and $\beta=6/g^2$ is the gauge coupling. $M$ is the quark matrix and $\kappa$ is the hopping parameter corresponding to the inverse of the quark mass for a heavy quark.
Applying the reweighting technique, we discuss the $(\beta, \kappa, \mu)$-dependence of the distribution function,
\begin{eqnarray}
  \frac{w(P; \beta, \kappa, \mu)}{w(P; \beta, \kappa_0, 0) }
  = \frac{\int {\cal D} U \delta(\hat{P}-P) 
  [\det M(\kappa, \mu)]^{N_{\rm f}} e^{6 \beta N_{\rm site} \hat{P}}
  }{\int {\cal D}U \delta(\hat{P}-P) [\det M(\kappa_0, 0)]^{N_{\rm f}}
  e^{6 \beta_0 N_{\rm site} \hat{P}} }
  = e^{6 (\beta - \beta_0) N_{\rm site} P}
  \left\langle \left[ \frac{ \det M(\kappa, \mu)}{\det M(\kappa_0, 0)} 
  \right]^{N_{\rm f}}\right\rangle_P ,
\label{eq:R}
\end{eqnarray}
where $\langle \cdots \rangle_P$ is the expectation value with fixed 
the plaquette $P$ measured in a simulation at $(\kappa_0, \mu=0)$. 
We now define the effective potential as
$ V_{\rm eff}(P; \beta, \kappa, \mu) = -\ln w(P; \beta, \kappa, \mu). $
The $(\beta, \kappa, \mu)$-dependence is given by
\begin{eqnarray}
V_{\rm eff}(P; \beta, \kappa, \mu) 
= V_{\rm eff}(P; \beta_0, \kappa_0, 0) - 6 N_{\rm site} (\beta-\beta_0)\, P 
- \ln \left\langle \left[ \frac{ \det M(\kappa, \mu)}{\det M(\kappa_0, 0)} 
  \right]^{N_{\rm f}}\right\rangle_P .
\label{eq:betaR}
\end{eqnarray}

We study the heavy quark region where the hopping parameter $\kappa$
is sufficiently small, performing simulations at $\kappa_0=0$.
At the lowest order of the hopping parameter expansion, 
the quark determinants 
is evaluated as
\begin{equation}
  \frac{\det M(\kappa, \mu)}{\det M(0,0)} 
  = \exp \left[ 288N_{\rm site} \kappa^4 P +3 \times2^{N_t+2} N_s^3 
  \kappa^{N_t} \cosh \left( \frac{\mu}{T} \right) \left\{ \Omega_{\rm R}
  +i\tanh \left( \frac{\mu}{T} \right) \Omega_{\rm I} \right\} \right] ,
  \label{eq:detM}
\end{equation}
where $\Omega_{\rm R}$ and $\Omega_{\rm I}$ are the real and imaginary parts 
of the Polyakov loop. 
The first term proportional to $P$ can be absorbed into the gauge action 
by a shift $\beta \rightarrow \beta + 48N_{\rm f}\kappa^4$.
Because the term from the gauge action in Eq.~(\ref{eq:betaR}) is linear, 
this term does not contribute to the appearance of the critical point 
at which the second derivative of $V_{\rm eff}$ with respect to $P$ vanishes.
The complex phase of $\det M$ is induced by $\Omega_{\rm I}$, which causes the sign problem. 

We perform simulations of SU(3) pure gauge theory at five $\beta$ values 
in the range $5.68$ -- $5.70$ on a $24^3\times 4$ lattice.
The details of the simulations are given in Ref.~\cite{saito11}.
We combine the data obtained at five $\beta$ to enlarge the range of $P$.
Measuring the histogram of $P$, i.e. the plaquette distribution function 
Eq.~(\ref{eq:pdist}), we calculate the derivative of the effective potential.
The $\beta$-dependence of $dV_{\rm eff}/dP$ at fixed $\kappa$ is controlled by 
$dV_{\rm eff}/dP(\beta)
=dV_{\rm eff}/dP(\beta_0)-6N_{\rm site}(\beta-\beta_0)$.
Since the additional reweighting term in this equation is constant, 
the shape of $dV_{\rm eff}/dP$ as a function of $P$ does not change under a change of $\beta$.
Therefore, $dV_{\rm eff}/dP$ is a good quantity to identify the order of phase transitions. 
If the transition is of first order, the effective potential has two minima. 
Then, $dV_{\rm eff}/dP$ must be an S-shaped function and crosses 
the horizontal axis at three values of $P$.

\begin{figure}[t] 
   \begin{minipage}{7.2cm}
   \includegraphics[width=7.2cm]{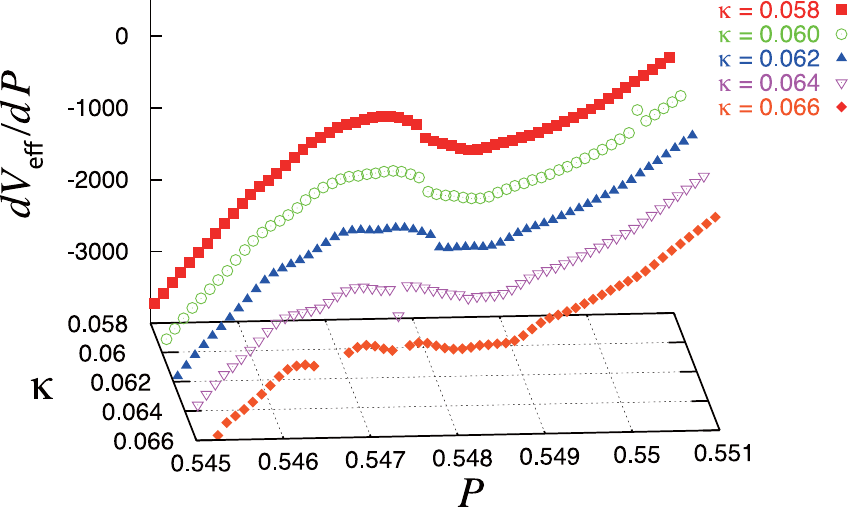} 
   \vspace{-8mm}
   \caption{Derivative of $V_{\rm eff}$ at nonzero $\kappa$ for $N_{\rm f}=2$.}
   \label{fig:dvdp_3d}
   \end{minipage}
   \hspace{0.8cm}
   \begin{minipage}{6.2cm}
   \includegraphics[width=6.2cm]{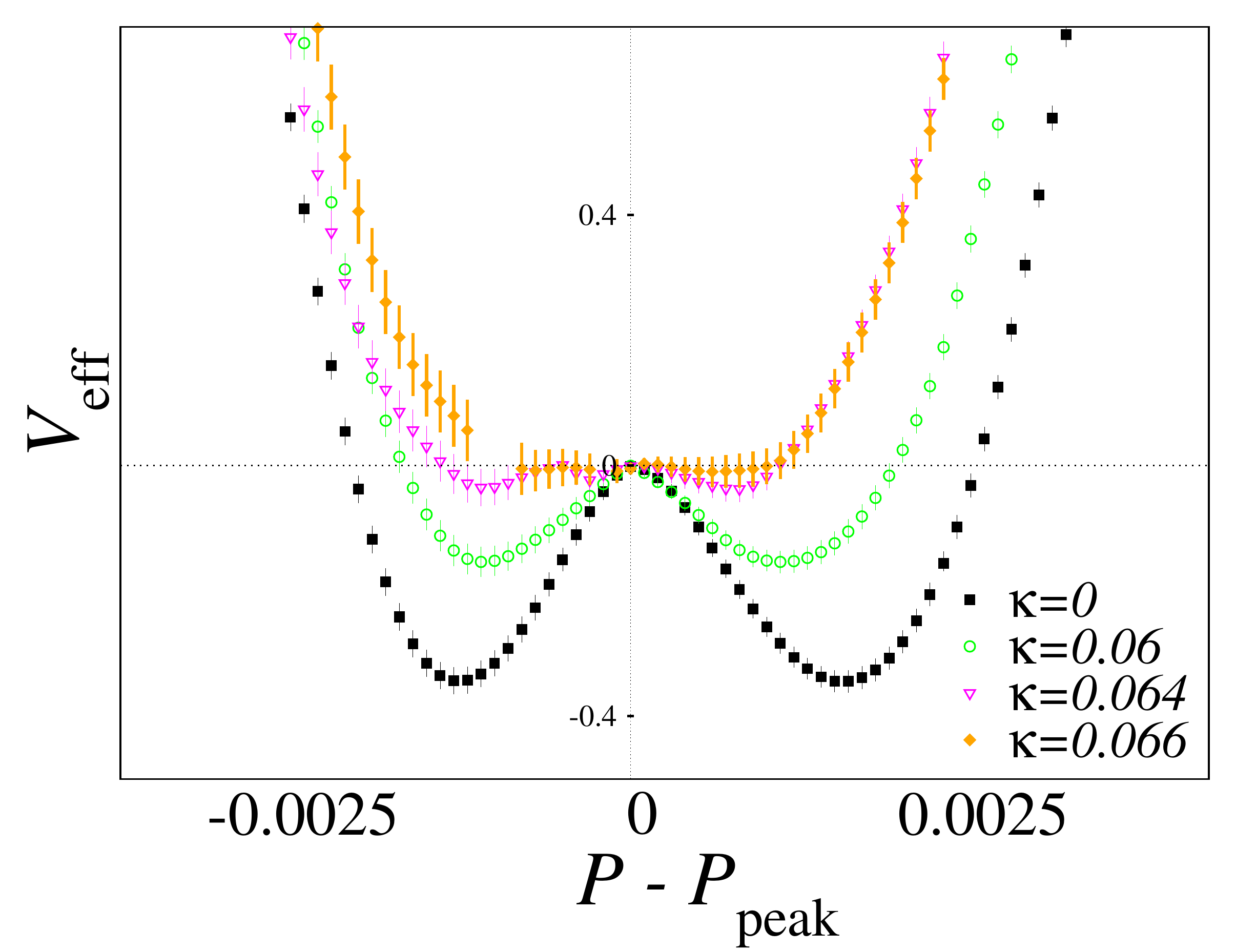}
   \vspace{-8mm}
   \caption{Effevtive potential at $\beta_c$ for each $\kappa$.}
   \label{fig:veff}
   \end{minipage}
\end{figure}

The $\kappa$-dependence of the effective potential is investigated at $\mu=0$ by the reweighting method
up to the order $\kappa^4$.
Using the data of $w(P;\beta, \kappa=0, \mu=0)$ and reweighting it, we evaluate $dV_{\rm eff}/dP$ at nonzero $\kappa$. 
Results near the critical $\kappa$ for $N_{\rm f}=2$ are plotted 
in Fig.~\ref{fig:dvdp_3d}. 
The S-shape structure becomes weaker as $\kappa$ increases 
and turns into a monotonically increasing function around $\kappa = 0.066$. 
This behavior suggests that the first order phase transition at $\kappa=0$
becomes a crossover at $\kappa \approx 0.066$. 
We also show, in Fig.~\ref{fig:veff}, the $\kappa$-dependence of 
$V_{\rm eff}$ at the transition point obtained by a numerical integration 
of $d V_{\rm eff}/ dP$. 
The double-well becomes shallower as $\kappa$ increases.

\begin{figure}[t] 
   \begin{minipage}{6.3cm}
   \includegraphics[width=6.3cm]{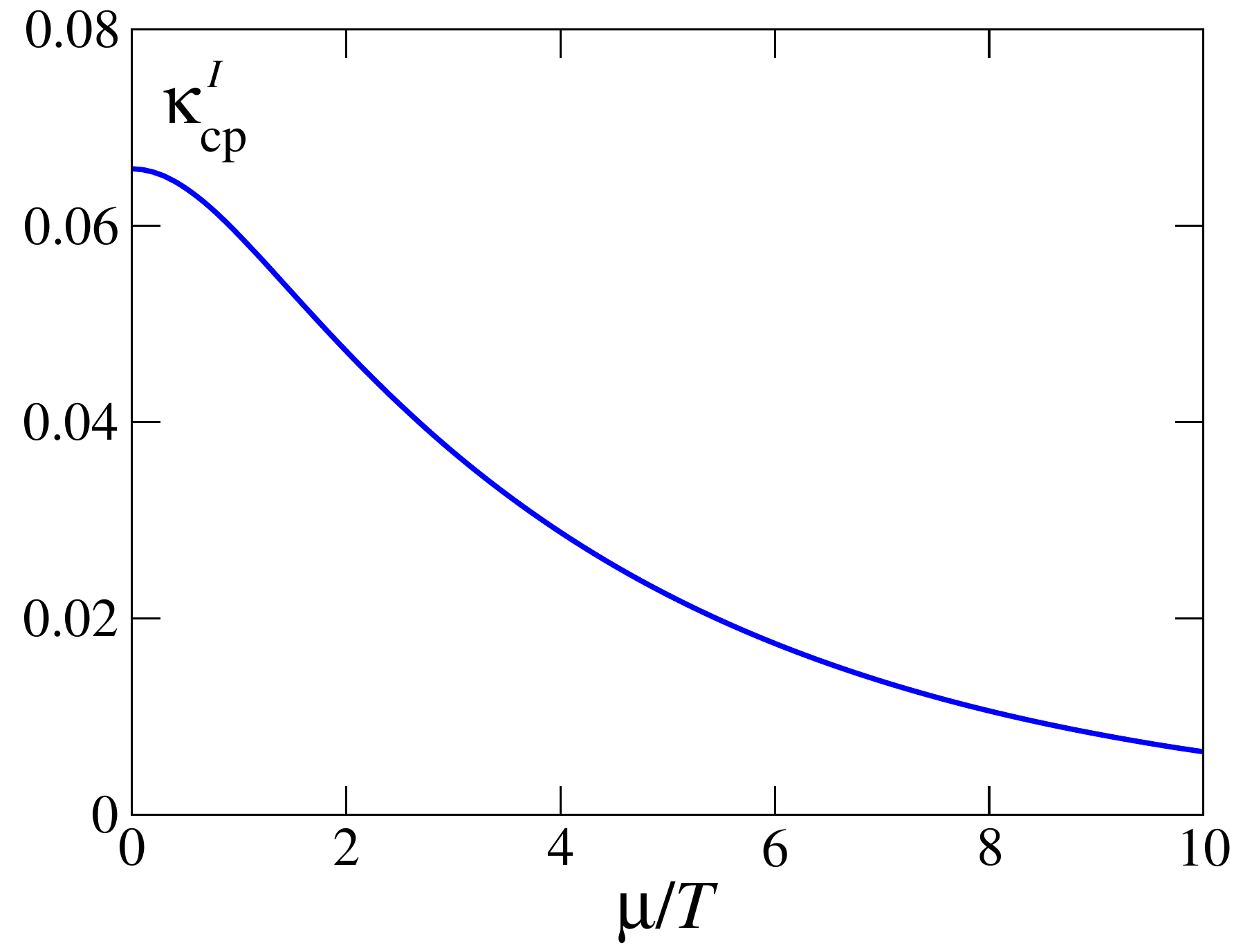} 
   \vspace{-8mm}
   \caption{Critical point in the phase-quenched approximation 
   for $N_{\rm f}=2$.}
   \label{fig:kcp_finitemu}
   \end{minipage}
   \hspace{0.8cm}
   \begin{minipage}{6.3cm}
   \includegraphics[width=5.0cm]{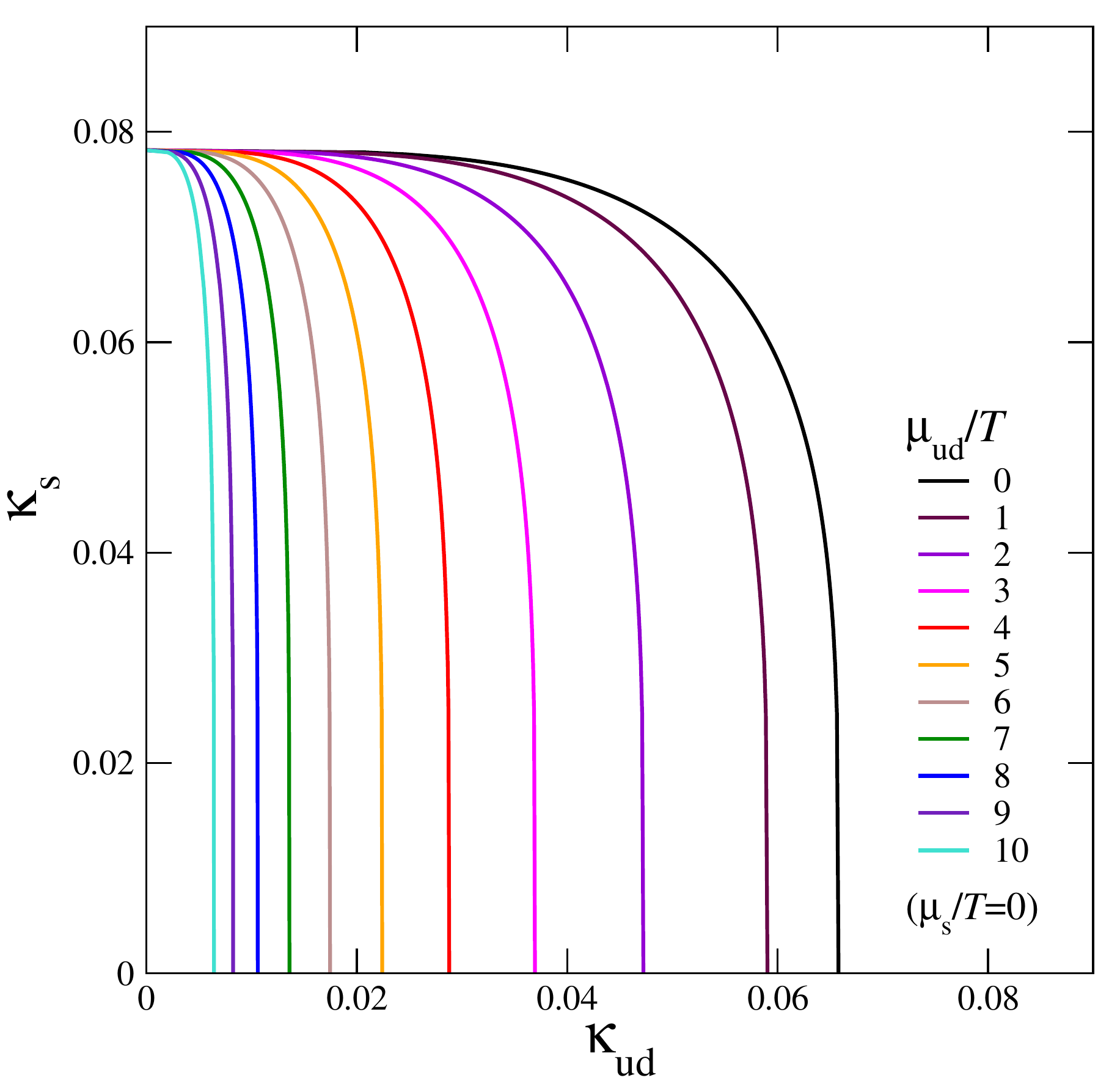}
   \vspace{-6mm}
   \caption{Critical line for each $\mu_{\rm ud}/T$ in $N_{\rm f}=2+1$ QCD
            with $\mu_{\rm s}/T=0$.}
      \label{fig:kcp_2+1}
   \end{minipage}
\end{figure}

We next consider the case of phase quenched finite density QCD, in which the complex phase term is neglacted in Eq.~(\ref{eq:detM}).
In 2-flavor QCD, this corresponds to the case of the isospin chemical potential, $\mu_u=-\mu_d \equiv \mu$. 
We find that, to the lowest order of the hopping parameter expansion, 
the effect of $\mu$ is just to modify the hopping parameter as
$\kappa \rightarrow \kappa\, \cosh^{1/N_t} (\mu/T)$ in the theory at $\mu=0$.
Therefore, the critical point at nonzero $\mu$ is given by
$\kappa_{\rm cp}^I(\mu)=\kappa_{\rm cp}(0) / \cosh^{1/N_t}(\mu/T),$
where $\kappa_{\rm cp}(0)$ is the critical point at $\mu=0$. 
We plot the critical line in Fig.~\ref{fig:kcp_finitemu}. 
The critical value decreases exponentially as $\mu/T$ increases.

The complex phase of $\det M$ on the critical line is given by 
$\theta =3 \times 2^{N_t+2} N_s^3 \kappa_{\rm cp}^{N_t}(0) 
\tanh(\mu/T) \ \Omega_{\rm I}$.
Because $\tanh(\mu/T)<1$, the upper limit of the complex phase fluctuation can be estimated from the data of $\Omega_{\rm I}$ in the whole range of $\mu/T$.
It is found in Ref.~\cite{saitolat11} that the contribution from the complex phase to the location of the critical point is quite small in the analysis by quenched simulations on a $24^3 \times 4$ lattice.

Adopting the phase quenched approximation, 
the application to $N_{\rm f}=2+1$ is straightforward
because the difference from $N_{\rm f}=2$ is just the replacement of $2\kappa^{N_t}$ by $2 \kappa_{\rm ud}^{N_t} + \kappa_{\rm s}^{N_t}$. 
We then find the critical $(\kappa_{\rm ud}, \kappa_{\rm s})$ satisfy 
\begin{equation}
2 \kappa_{\rm ud}^{N_t}(\mu) \cosh(\mu_{\rm ud}/T) 
+ \kappa_{\rm s}^{N_t}(\mu) \cosh(\mu_{\rm s}/T) 
= 2 [\kappa_{\rm cp}^{N_{\rm f}=2}(0) ]^{N_t} .
\end{equation}
The critical lines in the $\kappa$ plane for up, down and strange are drown 
in Fig.~\ref{fig:kcp_2+1} 
for $\mu_{\rm ud}/T=0$ -- $10$ and $\mu_{\rm s}/T=0$.

\section{Effective potential at $\mu \neq 0$ by phase quenched simulations}
\label{sec:light}

Toward a study of the critical line in the light quark region,
we propose a method based on the histogram method combined with the cumulant expansion 
of the complex phase of $\det M$ \cite{nakagawa11}.
To handle the $\kappa$ and $\mu$ dependence in a wide parameter range, 
we study the probability distribution function for the plaquette $\hat{P}$ 
and the quark determinant 
$\hat{F}(\mu) \equiv N_{\rm f} \ln |\det M(\mu)/\det M(0)|.$
The combination of phase quenched simulations and the reweighting method is applied to evaluate the probability distribution function.
Decomposing the quark determinant as
$(\det M(\mu))^{N_{\rm f}} = e^{i\theta(\mu)}|\det M(\mu)|^{N_{\rm f}}$,
the distribution function $w(P,F; \beta, \mu)$ is calculated by 
\begin{equation}\label{eq:part_func}
w(P,F;\beta,\mu)
= \int {\cal D} U \delta(\hat{P}-P) \delta(\hat{F}-F) e^{i\theta(\mu)} 
|\det M(\mu)|^{N_{\rm f}} e^{6\beta N_{\rm site} \hat{P}}
= \left\langle e^{i\theta(\mu)} \right\rangle_{(P,F)}
w_0(P,F,\beta,\mu),
\end{equation}
where $w_0(P,F; \beta, \mu)$ is the distribution function in the phase quenched simulation and $\langle e^{i\theta} \rangle_{(P,F)}$ is the average of the phase factor with fixed $P$ and $F$ in the phase quenched simulation.
Introducing the phase quenched effective potential $V_0 = - \ln w_0$,
the effective potential $V_{\rm eff}$ can be written as
$V_{\rm eff} = V_0 - \ln \langle e^{i\theta} \rangle.$

\begin{figure}[t] 
   \begin{minipage}{6.3cm}
   \vspace*{-2mm}
   \includegraphics[width=6.3cm]{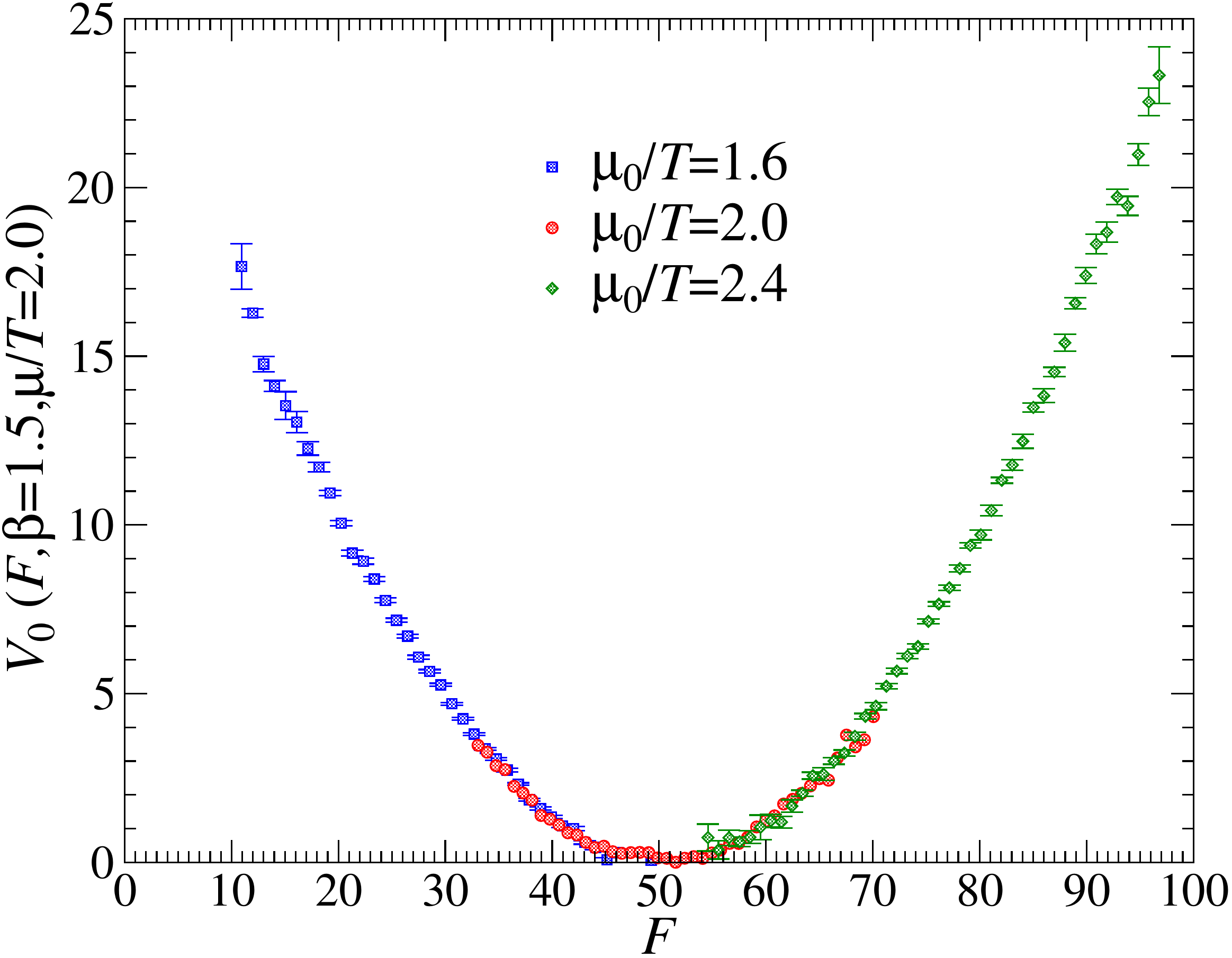}
   \vspace{-8mm}
   \caption{Effective potential $V_0$ at $\mu/T=2.0$ evaluated at 3 simulation points.}
   \label{fig:veff_pqs}
   \end{minipage}
   \hspace{0.8cm}
   \begin{minipage}{6.3cm}
   \includegraphics[width=6.3cm]{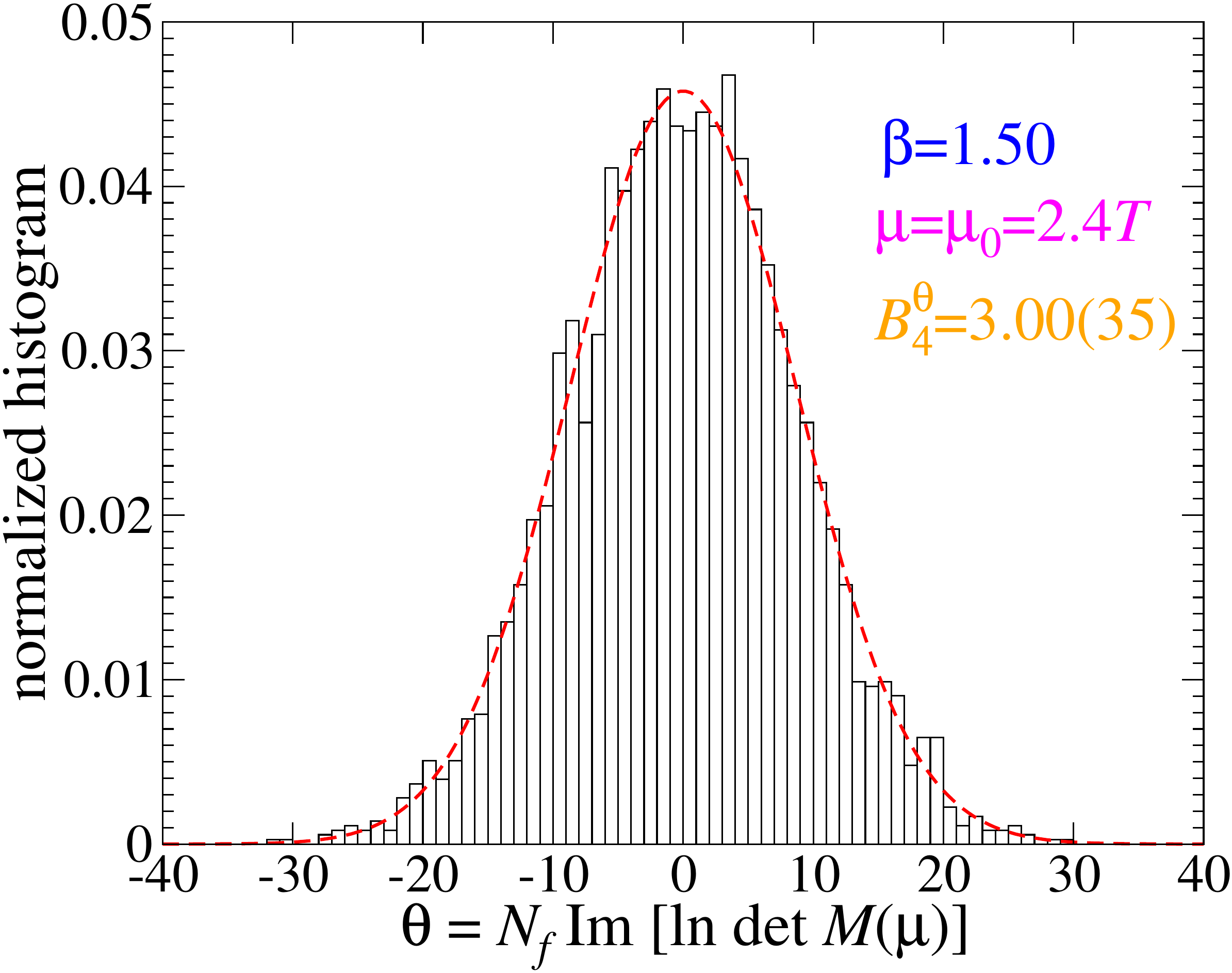} 
   \vspace{-8mm}
   \caption{Distribution of the complex phase of the quark determinant.}
   \label{fig:theta}
   \end{minipage}
\end{figure}

In the study,
there are two important problems.
The first problem is that the effective potential in a wide range of $(P, F)$ parameter plane is needed to investigate the shape of the effective potential. 
As is done in the previous section, we can enlarge the parameter range combining data obtained by different simulations.
We perform simulations at several $\mu$ and combine these data using the reweighing method.
$V_0$ at $\mu$ is evaluated from the data at $\mu_0$ by
$V_0(\mu)=V_0(\mu_0)- \ln R(\mu,\mu_0),$ 
where $R=w_0(\mu)/w_0(\mu_0) 
= \langle |\det M(\mu)/ \det M(\mu_0)|\rangle_{(P,F)}$.
Using this equation, we plot $V_0$ at $\mu/T=2.4$ in Fig.~\ref{fig:veff_pqs}.
The data are taken with the RG-improved Iwasaki gauge action and $N_{\rm f} =2$ clover improved Wilson quark action on an $8^3 \times 4$ lattice.
The data obtained at different simulation point are consistent.
Combining these data, we evaluate $V_0$ in a wide range of $F$.

The next problem is the sign problem.
If the sign of $e^{i \theta}$ changes frequently, the error of 
$\langle e^{i \theta} \rangle_{(P,F)}$ becomes larger than 
the mean value.
To aboid this problem, we calculate $\langle e^{i \theta} \rangle_{(P,F)}$ 
by a cumulant expansion,
$\langle e^{i\theta(\mu)} \rangle_{(P,F)}
= \exp [ i\langle \theta \rangle_c
- \langle \theta^2 \rangle_c /2
- i\langle \theta^3 \rangle_c /3!
+ \langle \theta^4 \rangle_c /4!
+ \cdots ],$
where $\langle \theta^n \rangle_c$ is the $n$-th order cumulant of $\theta$.
In this equation, the odd order cumulants cause the sign problem, but
these terms must be zero from the symmetry under $\mu$ to $-\mu$.
Once we remove the odd terms, the sign problem changes to the convergence problem of the cumulant expansion.

When $\theta$ is small, this expansion converges. 
However, if 
the distribution of $\theta$ is almost uniform 
in $-\pi < \theta \leq \pi$, the expansion dose not converge.
We adopt an alternative definition of the phase in the range 
$-\infty < \theta < \infty$.
We measure the derivatives of $\ln\det M$ with respect to $\mu$,
and define the phase by integrating the derivatives over $\mu$,
\begin{equation}\label{eq:mu_integral_theta}
\theta(\mu) = N_{\rm f} {\rm Im} \left[ \ln \det M(\mu) \right]
            = N_{\rm f} \int^{\mu/T}_0 {\rm Im} \left[
              \frac{\partial(\ln\det M(\mu))}{\partial(\mu/T)}
              \right]_{\bar{\mu}} d\left( \frac{\bar{\mu}}{T} \right).
\end{equation}
We plot the distribution of $\theta$ in Fig.~\ref{fig:theta}.
The distribution is well approximated by a Gaussian function. 
The dashed line is the fitted Gaussian function. 
For such a case, the cumulant expansion converges even when the fluctuation of $\theta$ is large, since 
$\langle e^{i\theta} \rangle$ is given only by the second order cumulant
if the distribution is Gaussian exactly.
Therefore, the result of the $\theta$ distribution suggests that the estimation 
of the complex phase factor by the cumulant expansion works well even 
in the high density region where the complex phase fluctuation is rather large. 
The details of the results shown in this section are given 
in Ref.~\cite{nakagawa11}.

\section{Summary}
\label{sec:summary}

We studied the QCD phase diagram in the heavy quark region.
The critical surface which separates the first order region
and the crossover region in the $(\kappa_{\rm ud}, \kappa_{\rm s}, \mu)$ space 
is calculated performing simulations in the heavy quark limit.
We moreover discussed a new method to study the critical surface in the light quark region.

\end{document}